\documentstyle[preprint,aps,epsf]{revtex}

\newcommand{\beq}{\begin{eqnarray}}
\newcommand{\eeq}{\end{eqnarray}}

\newcommand{\eps}{\epsilon}

\newcommand{\e}{{\rm e}}
\newcommand{\cc}{{\rm c.c}}

\newcommand{\btem}{\bibitem}

\newcommand{\PRL}{Phys.\ Rev. \ Lett.}

\begin{document}
\preprint{RYUTHP-97/1, May 1997}

\draft

\title{Dynamical Reduction of Discrete Systems Based on the Renormalization 
 Group Method}

\author{T. Kunihiro and J. Matsukidaira}

\address{Faculty of Science and Technology, Ryukoku University,
 Seta, Ohtsu, 520-21, Japan}

\date{\today}
 
\maketitle

\begin{abstract}
The renormalization group (RG) method is
 extended for  global asymptotic analysis of discrete systems.
We  show that the RG equation in the discretized form 
 leads to difference equations corresponding to
 the Stuart-Landau or Ginzburg-Landau equations. We propose
a discretization scheme which leads to a faithful  discretization
of  the reduced dynamics of the original differential equations. 
\end{abstract}

\narrowtext

It is a fundamental problem in physics, especially,
 in statistical physics since Boltzmann,  
 to reduce many degrees of freedom of a dynamical system to fewer 
 degrees of freedom preserving the essential nature of the system.
\cite{holmes}. 
The reduced dynamics is often described  by 
 a few  collective variables which represent slow and long-wave length 
 motion of the system.  There are several methods for the 
dynamical reduction such as the multiple-scale methods (including the
 reductive perturbation method\cite{kuramoto,manneville}), 
the average methods (including Whitham
 method\cite{whitham}), the method of normal forms\cite{normal} and so on.
 We are  usually interested in asymptotic behavior
 of the system after a long time. 
 Thus the  problem is to  obtain a dynamical reduction
 in asymptotic and global domains. 

Recently it has been recognized and emphasized 
by  Chen, Goldenfeld and Oono
 \cite{oono} that the renormalization group equation 
first developed in quantum field theory\cite{rg}  is  a powerful 
tool for  global and asymptotic analysis: 
They applied the RG equation $\grave{a}$ la Gell-Mann-Low 
 to ordinary and partial differential equations and 
  showed that the RG equation nicely gives a reduction of the dynamics
 describing slow and long-wave length 
motions of the system; the reduced dynamics is 
 described by   so called the amplitude equations.
 Afterwards the reason of the powerfulness of the RG equations  
 was accounted for in the context of
 classical theory of envelopes\cite{kuni}.
 More recently, it has been shown that the RG method can also naturally 
leads to  the phase equations as well as the amplitude 
equations\cite{sasa}.

 The purpose of this paper is twofold:
One is to show that the RG method can be extended to {\em discrete} systems 
and  leads  to 
 dynamical reduction of {\em discrete} dynamical systems (maps).
  A discrete system is 
 described by a {\em difference} equation. We notice that the extension is
highly non-trivial because the applicability of the RG method for differential 
equations is essentially relied on the local nature of 
 differentiation\cite{kuni}.
 Another purpose is to propose a discretization scheme for differential
 equations:  It is needless to say that different discretization schemes of 
non-linear differential equations leads to different dynamical systems.
 We shall show that such a discretization-scheme dependence 
 also exists for the reducibility of the dynamics and give
a  good discretization scheme based on the notion of the reducibility of the
 dynamics.
 
Now let us take the following  discrete  system
 as a typical example of nonlinear discrete systems,
\beq
x_{n+2}-2\cos \theta x_{n+1}+x_n=\eps f(x_n,x_{n+1}),
\eeq
where $\theta$ is a constant and the function $f(x_n,x_{n+1})$
 contains nonlinear terms. We remark that the equation can be 
 converted to a vector equation,${\bf X}_n={\bf F}({\bf X}_n)$ 
 with ${\bf X}_n=\, ^t(x_n, y_n\equiv x_{n+1})$ and 
${\bf F}({\bf X}_n)=\,^t(y_n, 2\cos \theta y_n-x_n+\eps f(x_n, y_n))$.
 We have taken an example of second-order equation here, but the 
 following discussion is applicable for higher-order equations. 
 We notice that the unperturbed equation ($\eps=0$) has a neutrally
stable
 solution ${\rm exp}(\pm i \theta)$.\cite{comment2} 
 Assuming  that $\eps$ is small,
 we shall apply the perturbation theory.
To make the discussion definite, let us take a concrete form for 
 $f(x_n,x_{n+1})$ as;
 $f(x_n,x_{n+1})=a_1x_n+a_2x_n^2+a_3x_n^3$.
If we put $a_2=0$, the resultant equation may be considered as a
discrete
 version of the  damped Duffin equation.

Expanding $x_n$ as
 $x_n=x_n^{(0)}+\eps x_n^{(1)}+\eps^2x_n^{(2)}+ \cdots$, let us
 try to find a solution which is valid around $n=n_0$ where $n_0$ 
 is arbitrary: $x^{(i)}_n\ (i=0, 1, ...)$ satisfies 
$\hat{L}x_n^{(0)}=0,$ 
$\hat{L}x_n^{(1)}= a_1x_n^{(0)}+a_2{x_n^{(0)}}^2+a_3{x_n^{(0)}}^3,$ 
and so on. Here $\hat{L}=E^2 -2\cos \theta \cdot E+ 1$ with $E$ is 
 the forwarding operator, i.e., $Ex_n=x_{n+1}$. 
 The lowest order solution may be written as 
\beq
x_n^{(0)}=A(n_0)\e ^{in\theta}+ \cc.
\eeq
Here we have made it explicit that $A$ may be   a function of $n_0$;
 its functional dependence is yet unknown and will be determined by 
 the RG equation, which determination constitutes the central part of 
the RG method.
The first order correction $x_n^{(1)}$ is given by 
\beq
x_n^{(1)}=-\frac{i}{2\sin \theta}(n-n_0)(a_1A+3a_3\vert A\vert^2 A)
           \e^{i(n-1)\theta} +\cc ,
\eeq
where we have omitted non-singular terms which are proportional
 to ${\rm exp}(ikn\theta)$ only with $k=0,2,3$. Notice that there
 appeared a secular term proportional to $n-n_0$. We remark that 
 the solution of the first-order equation is not unique; one could add
 any term proportional to $x_n^{(0)}$. We have chosen the above form so 
 that only  new independent terms appear and the secular term vanishes
 at $n=n_0$, which assures  the lowest order approximation is as good
 as possible. 
Up to $O(\eps^2)$,we have an approximate solution 
 $x_n=x_n^{(0)}+\eps x_n^{(1)}\equiv x_n((A(n_0);n_0)$, which is only 
locally valid around $n=\forall n_0$;
 the validity as an approximate solution is lost due to 
the secular term as $\vert n-n_0\vert$ becomes large. 
It means that the naive perturbation expansion breaks, 
which is a well known fact.

One may say, however, that  we have a family of discrete curves 
$x_n(A(n_0), n_0)$
 with $n_0$ being a parameter characterizing the curves\cite{kuni}.
 Each curve will become a good approximation for the exact solution  
around $n=n_0$ if $A(n_0)$ is suitably chosen.  Therefore the 
``envelope" 
of the  family of the curves may give a good approximation of the exact 
solution  in a global domain.\cite{comment1}
The ``envelope" of $x_n((A(n_0);n_0)$ is constructed as follows: 
We first impose that
\beq
\Delta_{n_0}x_n(A(n_0);n_0) =0,
\eeq
 where $\Delta_{n_0}$ denotes the difference w.r.t. $n_0$. 
 This is the basic equation of our method and we call it the
 RG/E equation.
 This is an equation to give $n_0$  and $A(n_0)$
 as  functions of $n_0(n)$. The ``envelope'' $x_n^{\rm E}$ is given by 
inserting this solution to $x_n(A(n_0);n_0)$;
 $x_n^{\rm E}=x_n(A(n_0(n));n_0(n))$.
 However, since we are
 constructing the ``envelope'' which contacts with the local solutions
 at $n=n_0$ so that the ``envelope'' give a good approximation, $n_0$ 
 should be $n$,i.e., $n_0=n$: Notice that this choice
 nicely eliminates the secular term from  $x_n^{\rm E}$.
 Conversely speaking, $A(n)$ can be 
 determined so that Eq.(4) gives $n_0=n$; this possibility is related with 
 the ``renormalizability'' of the equation\cite{oono}. 
 Thus we have 
\beq
\Delta A(n)=\eps \frac{{\rm exp}(-i\theta)}{2i\sin \theta}
 (a_1A(n)+3a_3\vert A(n)\vert ^2A(n)),
\eeq
where $\Delta$ denotes the difference operator w.r.t. $n$\cite{comment3}.
This is the amplitude equation in a discrete form corresponding to
Stuart-Landau equation \cite{landau} for continuum systems. 
 $x_n^{\rm E}$ is thus  given by
\beq
x_n^{\rm E}=(A(n)e^{in\theta} +\cc) +
2a_2\eps\vert A(n)\vert ^2+ {\rm h.h.}.
\eeq
here h.h. denotes the higher harmonics.

 A significant point is that this function gives an approximate but
uniformly valid  solution in a global domain.
Let us show this in a general setting. 
Let ${\bf X}_n=\, ^t(X_{1n}, X_{2n}, \cdots , X_{dn})$ and 
 ${\bf F}({\bf X}_n, n) =\, ^t(F _1({\bf X}_n, n)$, 
$F _2({\bf X}_n, n),\cdots , F_d({\bf X}_n, n))$; we assume that 
${\bf F}({\bf X}_n, n)$ is analytic with respect to ${\bf X}_n$.
We consider  the difference equation
\beq
\Delta {\bf X}_n = {\bf F}({\bf X}_n , n).
\eeq
We suppose that $\tilde{{\bf X}}_n({\bf W}(n_0), n_0)$ is 
an approximate solution to the equation up to  $O(\eps ^p)$,
where the $d$-dimensional vector ${\bf W}(n_0)$ denotes the initial values 
assigned at the initial time $n=n_0$.  Here notice that $n_0$ is arbitrary.
The envelope function is given by 
${\bf X}^{E}_n\equiv \tilde{\bf X}_n ({\bf W}(n),n),$
where ${\bf W}(n)$ is the solution to the RG/E equation,
$\Delta_{n_0}\tilde{\bf X}_n({\bf W}(n_0), n_0)=O(\eps^p)$.
Then one can show that {\em  ${\bf X}^{E}_n$ satisfies the original
 equation uniformly up to $O(\eps ^p)$} as follows:
$\Delta{\bf X}^{E}_n=\Delta \tilde{\bf X}_n({\bf W}(n_0), n_0)\vert_{n_0=n+1}
+\Delta_{n_0}\tilde{\bf X}_n({\bf W}(n_0), n_0)\vert_{n_0=n}$
$= {\bf F}(\tilde{\bf X}_n({\bf W}(n_0,n_0), n)\vert_{n_0=n+1} + O(\eps^p)$
 $= {\bf F}(\tilde{\bf X}_n({\bf W}(n),n), n) -\partial {\bf F}
/\partial \tilde{\bf X}_n\cdot\Delta_{n_0} \tilde{\bf X}_n({\bf W}(n_0), n_0)
\vert_{n_0=n}+O(\eps^{m\geq p})$
 $= {\bf F}({\bf X}^{E}_n,n)+O(\eps^p).$
 Here we have utilized the RG/E equation in the second and last equality,
and the analyticity of ${\bf F}$ in the third equality. 
This concludes the proof.

We notice that the amplitude equation Eq.(5) is 
 a first-order equation and a dynamical reduction is achieved in 
 comparison with the original equation. In fact, in the 
 polar representation $A(n)=R_n{\rm exp}(i\varphi_n)$, 
$R_n$ and $\varphi_n$
 satisfy the following equations,
\beq
R_{n+1}=R_n- \eps/2\cdot (a_1R_n+3a_3R_n^3),
\eeq
 and $\varphi _{n+1}=\varphi_n-\eps/2\cdot \cot \theta (a_1+ 3a_3R_n^2),$
 respectively. Notice that  $R_n$ 
 is determined  by the first order equation independently of 
$\varphi_n$, which in turn is given in terms of $R_n$.
Thus one may say that the second-order dynamical system Eq.(1) is 
reduced to a first-order one. 
As is seen from the derivation, this reduced equation has a universal 
nature as  has Stuart-Landau equation.

The first order equation for $R_n$ has  simple qualitative properties 
depending on the signs and values of $a_1$ and $a_3$.
  For example, when 
 $a_1<0$ but $a_3>0$, the equation is converted to 
$f_{n+1}=f_n+af_n(1-f_n^2)$ 
with $f_n=\sqrt{3a_3/\vert a_1\vert}R_n$ and 
$a\equiv \eps \vert a_1\vert/2$.
 For $0<a<1$, the equation has a fixed point $f^{\ast}=1$, while
 for $1<a\leq 1.246$, the map shows a two-period behavior, and after 
 that the map rapidly shows multiple-period behavior then eventually
 becomes chaotic. 

As another example of ordinary difference equation, let us take the 
 one which is derived as a discretization of the Rayleigh equation:
$\ddot{x} + x=\eps \dot{x}(1 -1/3\cdot \dot{x}^2).$
We remark that the
 equation admit a limit cycle with the radius of 2\cite{bender}.
 We take the following discretization;
$\ddot{x}\rightarrow  (x_{n+1}-2x_n+x_{n-1})/{\Delta t}^2 ,$
$\dot{x} \rightarrow  (x_n -x_{n-1})/\Delta t.$
That is, the central difference for the second derivative and the 
backward difference for the first derivative.   Thus we have
\beq
x_{n+1}-2\cos \theta x_n+ x_{n-1}=
\bar{\eps}(x_n-x_{n-1})
( 1- \frac{1}{3}\cdot \frac{(x_n-x_{n-1})^2}{{\Delta t}^2}),
\eeq
where  $\cos \theta =1-{\Delta t}^2/2$ and $\bar{\eps}=\eps \Delta t$.
We remark that this difference equation has a neutrally stable solution 
 as the unperturbed one.  This is not the case for other
 discretization schemes such as 
$\ddot{x}\rightarrow (x_{n+2}-2x_{n+1}+x_{n})/{\Delta t}^2$ and
$\dot{x} \rightarrow (x_{n+1} -x_{n})/\Delta t$. 
We  put $1/\Delta t=\omega$.

In the first-order approximation with respect to $\bar{\eps}$, 
we have
\beq
x_n&\equiv &x_n(A(n_0); n_0)=
       [A(n_0)+\bar{\eps}\frac{{\rm exp}(-i\theta)}{2i\sin \theta}
      (1-\omega^2\vert B(n_0)\vert ^2)B(n_0)\cdot (n-n_0)]e^{in\theta} 
    \nonumber \\ 
\ \ \ &  -&\frac{\bar{\eps}}{3}\omega^2
     \frac{B(n_0)^3{\rm exp}(3in\theta)}
 {({\rm exp}(3i\theta) -{\rm exp}(i\theta)) 
 ({\rm exp}(3i\theta) -{\rm exp}(-i\theta))}+{\rm c.c},
\eeq
where $B(n_0)=({\rm exp}(i\theta)-1)A(n_0)$.
Now the RG/E equation 
$\Delta _{n_0}x_n(A(n_0); n_0)\biggl\vert _{n_0=n}=0,$ 
gives the amplitude equation
$\Delta A(n)=\bar{\eps}{\rm exp}(-i\theta)/2i\sin \theta\cdot 
             (1-\omega^2\vert B(n)\vert^2)B(n),$
accordingly,
\beq
A(n+1)=A(n) + z A(n)(1 -\vert A(n)\vert ^2),
\eeq 
 with $z\equiv \bar{\eps}{\rm exp}(-i\theta)/\cos \theta/2$.
If we take the polar representation 
$A(n)=R_n{\rm exp}(i\varphi _n)$, we have
$R_{n+1}=R_n+ \eps a'R_n(1-R_n^2),$
 and  $\varphi _{n+1}=\varphi_n +\bar{\eps}\sin \theta/2 \cdot (1 -R_n^2)$
 with $a'=\Delta t\cos \theta /(2\cos \theta/2)$. We note that 
$a'$ as a function of $\Delta t$ is a parabola-like shape and takes
   the maximum about .6 at $\Delta t\simeq .9$;
 it vanishes at $\Delta t=0 $ and $\Delta t\simeq 1.4$.
Thus we see that $R_n$ goes up monotonically to a fixed 
point 1.

With this $R_n$ and $\varphi_n$, the envelope $x_n^{\rm E}$ 
is given as
\beq
x_n^{\rm E}\equiv x_n(A(n); n)=
 2R_n\cos(n\theta +\varphi_n) +\frac{\bar{\eps}}{12}
 \frac{\tan \theta}{\cos^3\theta/2}R_n^3\sin\{3(n\theta +\varphi_n) -
 \frac{3}{2}\theta\},
\eeq
which shows that the radius of the limit cycle is 2, 
 irrespective of the choice of $\Delta t$ in accordance with
 the original Rayleigh equation. 
 We remark that it is not the case for  other discretization schemes as 
given below Eq.(9).

Fig.1a shows $x^E_n$ given by Eq.(12) and the envelope $2R_n$ 
together with the exact solution of Eq.(9) with $\eps=.4$ and 
$\Delta t =0.25$. 
One can see that the agreement is excellent in the global domain;
 notice that the results is
 obtained in the first order approximation. One can also see  that the
 amplitude $2R_n$ successfully describe  the slow motion of the system.
 The characteristic features of the
 system as a dynamical system may be more clearly seen in Fig.1b where the
 behaviors in the ``phase space'' are shown.
 The agreement is excellent again.\cite{comment6}

Finally,  we consider  partial difference 
 equations. As an example, we take  the difference 
 equation which is given by a discretization of the 
 1-dimensional Swift-Hohenberg equation\cite{swift};
$\partial_t \phi(x,t) =\eps (\phi(x,t)-\phi(x,t)^3) -
 (1+\partial _x^2)^2\phi(x,t).$ 
With the discretization 
$\partial _t\phi \rightarrow (\phi(n, m+1)-\phi(n,m))/\Delta t
 \equiv \Delta _m\phi(n,m)/\Delta t, $
 $\partial_x^2\phi \rightarrow (\phi(n+1,m) -2\phi(n,m)+ \phi(n-1,m))
/\Delta x^2\equiv \Delta _n^2\phi(n,m)/{\Delta x}^2,$
 we  have the following partial difference equation
\beq
\hat{\cal L}\phi(n,m)=\eps (\phi(n,m) -\phi(n,m)^3)
\eeq
where 
$\hat{\cal L}\equiv \Delta _m+r({\Delta x}^2+\Delta _n^2)^2$,
 with $r\equiv \Delta t/{\Delta x}^4$.
We shall show that the difference equation admit a dynamical
reduction giving an amplitude equation which is the analogue of 
 the time-dependent Ginzburg-Landau equation in the continuum theory.
 The reason why the dynamical reduction is possible in the RG method
 is that the difference equation is constructed so that the 
 unperturbed equation has  a neutrally stable solution.

Making the Taylor expansion 
 $\phi =\phi_0 +\eps \phi_1 +\eps^2 \phi_2 + \cdots$, we have 
equations in the successive order
$\hat{\cal L}\phi_0=0, \ \ \ \hat{\cal L}\phi_1=\phi_0 -\phi_0^3,$
 and so on.  
We consider an asymptotic solution at $m \rightarrow \infty$ and
 take the following neutrally stable solution as the 0-th 
 order one;
\beq
\phi_0(n, m)=A(n_0, m_0)e^{in\theta _x}+{\rm c.c},
\eeq
with $\theta _x=2 {\rm Sin}^{-1}
 \Delta x/2$. Here
 we have made it explicit that the amplitude $A$ may 
 depend in a yet unknown way on the initial time $m_0$ and point $n_0$.
 They will be determined by the RG/E equation.

Then the first-order equation now reads
\beq
\hat{\cal L}\phi_1 =\{(A- 3\vert A\vert ^2 A)e^{in\theta_x}-A^3
e^{3i n\theta_x}\}
 +{\rm c.c.}.
\eeq
A straightforward but somewhat tedious manipulation gives 
\beq
\phi_1(n,m)=&\Biggl[&\biggl[ \mu_1 (m-m_0)-
 \frac{\mu_2r^{-1}}{8\sin ^2\theta_x}\{
   n^{(2)}-n_0^{(2)}+ i e^{-i\theta_x}(n-n_0)\}
                   \biggl]\nonumber \\ 
\ \ \ & \ & \times (A- 3\vert A\vert^2 A)e^{in \theta_x} 
  -\frac{A^3e^{3i(n+2)\theta_x}}{r(e^{3i\theta_x}-e^{i\theta_x})^2
                    (e^{3i\theta_x}-e^{-i\theta_x})^2}
                                 \Biggl] +{\rm c.c.},
\eeq
 where $n^{(2)}=n(n-1)$ and  $\mu_1+\mu_2=1$.
Thus we have an approximate solution 
$\phi(n,m)=\phi_0(n,m)+\eps \phi_1(n,m)$
 which is valid only for $(n,m)$ around $n\sim n_0$ and $m\sim m_0$.

Now the RG/E equation $\Delta _{m_0}\phi\vert _{m_0=m}=0$
 and $\Delta _{n_0}\phi \vert _{n_0=n}=0$  gives 
$\Delta _m A(n, m)= \eps \mu_1(1-3\vert A \vert ^2)A,$
 and $\Delta _n^2 A(n, m)= -\frac{\eps \mu_2}{4r\sin ^2\theta_x}
(1-3\vert A \vert ^2)A,$
respectively. 
   Here we have utilized the fact $\Delta _m A= O(\eps)$ and
 $\Delta _n A= O(\eps)$, and neglected terms of $O(\eps^2)$.
 Thus noting that $\mu_1+\mu_2=1$, we reach the amplitude equation
 for the difference equation
\beq
\Delta _mA(n,m)=
 4r\sin ^2\theta _x\Delta _n^2 A(n,m)+ 
 \eps (1- 3\vert A(n,m)\vert^2)A(n,m).
\eeq
 This is precisely the discretized form of the time-dependent 
 Ginzburg-Landau equation:
$\partial _t A(x,t)=4\partial_x^2\phi(x,t)+
 \eps (1- 3\vert A(x,t)\vert^2)A(x,t).$

In summary, we have shown that the renormalization group 
 method  can be nicely extended  to discrete systems, and that 
the method is useful as a tool for global asymptotic analysis and 
 gives 
 dynamical reduction of discrete systems. We have emphasized that 
the method is applicable for systems which have  neutrally 
stable solution.
 It is to be remarked that this is also the case for equations
 which have unperturbed solutions on invariant stable and unstable manifolds.
We have also given a notion of the discretization scheme that 
 faithfully preserves the nature of the reduced dynamics 
 irrespective of the magnitude of $\Delta t$.
 Finally, we notice that  it is not trivial what discretization scheme 
 preserves the integrability of differential equations
 which admit soliton solutions\cite{soliton}. 
  It will be interesting to examine if our discretization scheme
 based on the RG method can have a relevance to
  soliton theories.

We thank Prof. M. Yamaguti and Prof. R. Hirota for their
 interest in this work.
This work is supported by the Grants-in-Aid of the Japanese 
 Ministry of Education, Science and Culture, No. 07304065
 and No. 08640396.

\newcommand{\NG}{N. \ Goldenfeld}
\newcommand{\YO}{Y.\ Oono}

\newpage
 {\large Figure Captions}\\ 

\noindent 
{\bf Fig.1a}\ \ \  The dots shows $x^E_n$ (Eq.(12)) v.s. $n\Delta t$
 while the thin line
 the envelope $2R_n$ for $\eps=.4$ and $\Delta t =0.25$.  
 The  bold line shows the  exact solution of Eq.(9).\\ 
\noindent
{\bf Fig.1b}\ \ \ 
The  behaviors in the ``phase space''; the vertical axis denotes
 the ``velocity'' $v_n\equiv (x_{n+1} - x_n)/\Delta t$, 
while the horizontal axis $x_n$. 
The dots shows our approximate solution whereas the solid line 
 the exact one. 

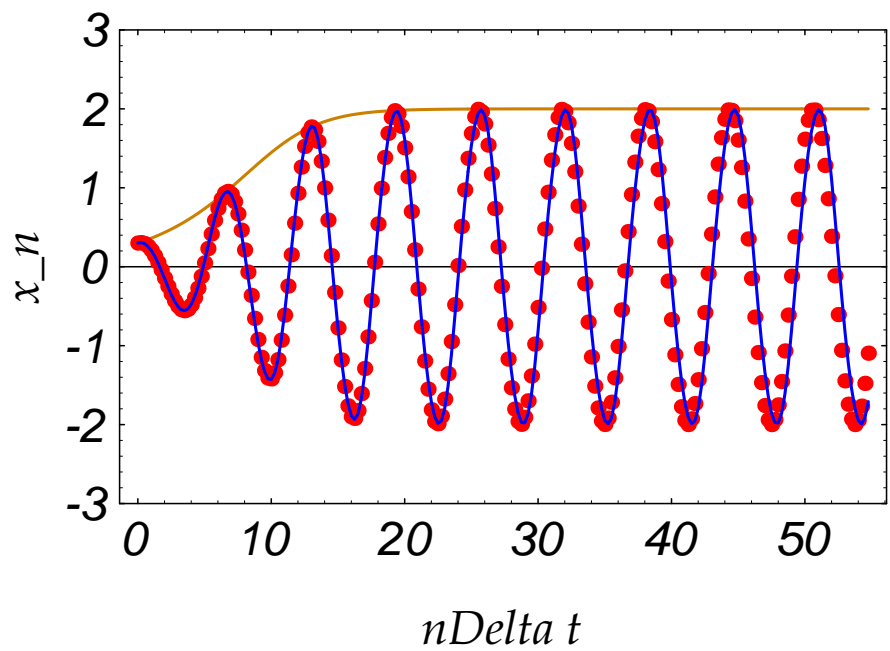
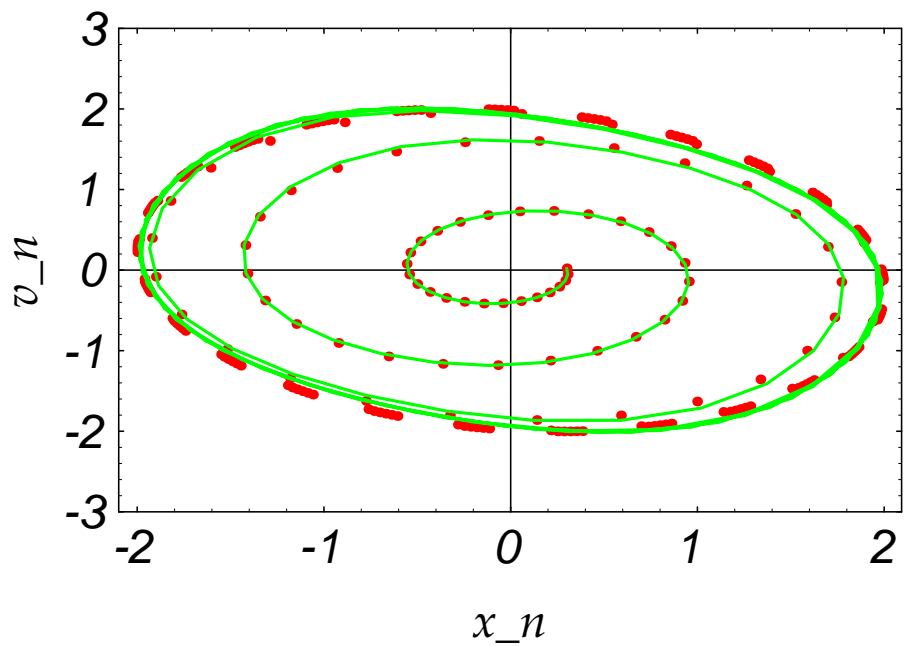
\end{document}